\documentclass[twocolumn,prd,aps,superscriptaddress,preprintnumbers,tightenlines,showpacs,nofootinbib,eqsecnum,amsfonts,amsmath, floatfix]{revtex4-1}
\usepackage{graphicx} 
\usepackage[usenames,dvipsnames]{xcolor}
\usepackage{xspace} 
\usepackage{bm} 
\usepackage[utf8]{inputenc} % for some references from inspirehep.net
\usepackage{eucal}
\usepackage{float}
\usepackage[normalem]{ulem}

\xdefinecolor{mylinkcolor}{rgb}{0,0,0.5}
\usepackage[
	bookmarksnumbered, 
	bookmarksopen, 
	bookmarksopenlevel=2,
	breaklinks=true,
	filecolor=mylinkcolor, 
	citecolor=mylinkcolor,
	linkcolor=mylinkcolor, 
	urlcolor=mylinkcolor, 
	menucolor=mylinkcolor,
]{hyperref}

\def\vct#1{{\bm{#1}}}

\allowdisplaybreaks

\newcommand{\nnm}{\nonumber}
\newcommand{\doe}{\partial}
\newcommand{\be}{\begin{equation}}
\newcommand{\ee}{\end{equation}}
\newcommand{\bse}{\begin{subequations}}
\newcommand{\ese}{\end{subequations}}

\newcommand{\mr}{\mathrm}

\newcommand{\mc}{\mathcal}

\newcommand{\bs}{\boldsymbol}

\newcommand{\AEI}{\affiliation{Max Planck Institute for Gravitational Physics (Albert Einstein Institute), Am M\"uhlenberg 1, Potsdam 14476, Germany}}
\newcommand{\Maryland}{\affiliation{Department of Physics, University of Maryland, College Park, MD 20742, USA}}

\begin{document}

\title{Energetics of two-body Hamiltonians in post-Minkowskian gravity
%\\ Alt: Energetics of effective-one-body Hamiltonians with post-Minkowskian information
% \\ Alt: Assessing the accuracy of post-Minkowskian effective-one-body Hamiltonians
% \\ Alt: Can the post-Minkowskian approximation improve inspiral waveforms?
}
\author{Andrea Antonelli}\AEI
\author{Alessandra Buonanno}\AEI\Maryland
\author{Jan Steinhoff}\AEI
\author{Maarten van de Meent}\AEI
\author{Justin Vines}\AEI

\date{\today}

\begin{abstract}

Advanced methods for computing perturbative, quantum-gravitational scattering amplitudes show great promise for improving our knowledge of classical gravitational dynamics.  This is especially true in the weak-field and arbitrary-speed (post-Minkowskian, PM) regime, where the conservative dynamics at 3PM order has been recently determined for the first time, via an amplitude calculation.  
 Such PM results are most relevantly applicable to relativistic scattering (unbound orbits), while bound/inspiraling binary systems, 
the most frequent sources of gravitational waves for the LIGO and Virgo detectors, are most suitably modeled by the weak-field and slow-motion (post-Newtonian, PN) approximation.  Nonetheless, it has been suggested that PM results can independently lead to improved modeling of bound binary dynamics, especially when taken as inputs for effective-one-body (EOB) models of inspiraling binaries. Here, we initiate a quantitative study of this possibility, by comparing PM, EOB and PN predictions for
	the binding energy of a two-body system on a quasi-circular
	inspiraling orbit against results of numerical relativity (NR) simulations. 
	The binding energy is one of the two central ingredients (the other being the gravitational-wave energy flux) that enters the computation of gravitational waveforms employed by LIGO and Virgo detectors, and for (quasi-)circular orbits it provides an accurate diagnostic of the conservative sector of a model. 
	We find that, whereas 
	3PM results do improve the agreement with NR with respect to 2PM (especially when used in the EOB framework), it is crucial to push 
	PM calculations at higher orders if one wants to achieve better performances than current waveform models used for LIGO/Virgo data analysis.  
\end{abstract}

\maketitle

\section{Introduction}
\label{sec:intro}

Gravitational waves (GWs) from binary black holes (BHs) and neutron
stars (NSs)~\cite{Abbott:2016blz,TheLIGOScientific:2016pea,TheLIGOScientific:2017qsa,LIGOScientific:2018mvr}
encode information about the structure of compact objects and their
interaction via (strong, dynamical) gravitational fields.  The
continuously improving network of GW detectors~\cite{TheLIGOScientific:2014jea,TheVirgo:2014hva,Aso:2013eba,LIGOIndia}
offers unprecedented insights into astrophysics and fundamental physics. Likewise, a continuous improvement in the accuracy of existing GW predictions, using both numerical and analytical methods, is necessary in order to continue the successful story of GW astronomy.

Regarding GW predictions for compact binary coalescence, numerical and
analytical methods complement each other well, since the analytic
post-Newtonian (PN, weak-field and slow-motion) approximation (e.g., see Refs~\cite{Blanchet:2013haa,Schafer:2018kuf,Futamase:2007zz,Poisson:2014,Goldberger:2007hy,Rothstein:2014sra}) 
relies on the separation between the orbit's and the body's scales
being large, while current numerical-relativity (NR) simulations (e.g., see Refs.~\cite{Mroue:2013xna,Chu:2015kft,Husa:2015iqa}) become inefficient in this regime.  Since
the orbital separation shrinks over time due to energy and angular
momentum loss via GW emission, a synergistic approach between both
methods is needed to predict the complete inspiral-merger-ringdown
(IMR) sequence for the compact binaries now routinely detected by
ground-based GW observatories~\cite{LIGOScientific:2018mvr}.

The effective-one-body (EOB) formalism~\cite{Buonanno:1998gg,Buonanno:2000ef} improves 
the accuracy of the (perturbative) PN two-body dynamics  (see, e.g., Refs.\cite{LeTiec:2011bk,Damour:2011fu,LeTiec:2011dp,Tiec:2013twa,Ossokine:2017dge}) by resumming PN results in such a 
way as to include the exact test-particle limit. EOB waveforms~\cite{Taracchini:2013rva,Pan:2013rra,Bohe:2016gbl,Nagar:2018zoe} are an important 
class of IMR waveform models employed in LIGO/Virgo searches and inference studies~\cite{
Abbott:2016blz,Abbott:2016nmj,TheLIGOScientific:2016pea,
TheLIGOScientific:2016wfe,TheLIGOScientific:2016src,Abbott:2016izl,TheLIGOScientific:2017qsa,LIGOScientific:2018mvr}. Because of 
the more accurate description of the dynamics toward merger (with respect to PN), 
EOB waveforms are also employed to build another class of IMR waveforms, the phenomenological waveform models 
(e.g., see Ref.~\cite{Khan:2015jqa}). In order to improve EOB waveform models in the entire binary's parameter space (i.e., large-mass ratios and large 
spins), better understand the uniqueness and robustness of the EOB resummation, and gain confidence in its range of
applicability, it is important to extend the EOB formalism to highly relativistic 
bound and unbound orbits. The large mass-ratio case, which is relevant for space-based and third-generation 
ground-based detectors and requires a very accurate modeling of fast-motion effects, is one important example~\cite{Yunes:2009ef,Damour:2009sm,Yunes:2010zj,Barausse:2011dq,Akcay:2012ea}, which we will follow up elsewhere~\cite{Antonelliea:2019}. Here, we focus on the post-Minkowskian (PM)
approximation (i.e., weak field and fast motion)~\cite{Blanchet:2013haa,Poisson:2014,Bertotti:1956,Bertotti:1960,Rosenblum:1978zr,Bel:1981be,Damour:1981bh,Portilla:1979xx,Portilla:1980uz,Westpfahl:1979gu,Westpfahl:1985,Schafer:1986,Westpfahl:1987}
applicable to scattering binaries (see also
Refs.~\cite{Ledvinka:2008tk,Foffa:2013gja,Damour:2016gwp,Bini:2017xzy,Vines:2017hyw,Damour:2017zjx,Bini:2018ywr,Blanchet:2018yvb,Bjerrum-Bohr:2013bxa,Holstein:2004dn,Bjerrum-Bohr:2018xdl,Neill:2013wsa,Vaidya:2014kza,Cheung:2018wkq,Kosower:2018adc}
for recent applications).

A crucial ingredient of the EOB formalism is the energy map \cite{Buonanno:1998gg} 
between the two-body and the effective one-body description.  While the
energy map was established as a natural choice up to 4PN order
\cite{Buonanno:1998gg,Damour:2000we,Damour:2015isa}, its properties become more 
apparent and unique (at least at 1PM) when extending the conservative EOB Hamiltonian to 1PM and 2PM 
orders \cite{Damour:2016gwp,Damour:2017zjx}.  One can also gain insight into EOB spin
maps at 1PM and 2PM orders~\cite{Vines:2017hyw,Vines:2018gqi,Bini:2018ywr}. These results, together with the prospect of
creating a waveform model for scattering binaries, certainly provide a
good motivation to push the PM knowledge to higher orders. Quite interestingly, 
profiting from recent advances in the area of scattering amplitudes
\cite{Parke:1986gb,Bern:1994zx,Bern:1994cg,Britto:2004nc,Britto:2004ap,Britto:2005fq,Bern:2008qj,Bern:2010yg,Bern:2010ue,
Guevara:2017csg,Arkani-Hamed:2017jhn,Ochirov:2018uyq}, the conservative Hamiltonian 
for a two-body system has been recently derived at 2PM order~\cite{Damour:2017zjx,Cheung:2018wkq} and 
3PM order~\cite{Bern:2019nnu}.  

However, the most frequent sources of GWs for LIGO/Virgo experiments
are bound/inspiraling binaries, instead of unbound/scattering ones.
It is not clear \textit{a priori} whether the source modeling 
of coalescing binaries for GW detectors will take real advantage of PM results, thus motivating 
to push PM calculations at even higher orders and extend them to the dissipative 
sector. It is also unclear whether 
insights on (and explicit resummations for) the EOB
Hamiltonian from PM results are already useful to improve the accuracy
of quasi-circular, inspiral waveforms for LIGO/Virgo analyses. Here, 
we start to shed light on these important inquiries by comparing PM, EOB and PN predictions for
the binding energy of a two-body system on a quasi-circular
inspiraling orbit against results of NR simulations. Indeed, the binding energy is 
one of the two central ingredients (the other being the GW energy flux) 
that enters the computation of gravitational waveforms (e.g., see Refs.~\cite{Buonanno:2009zt}). 
Thus, assessing the accuracy of PM predictions against the (``exact'') NR results, and quantifying the differences 
with respect to the EOB/PN results currently used in building waveform models for LIGO/Virgo analyses, is 
very relevant and timely, given also the strong interest that PM calculations have recently generated in the theoretical high-energy physics community.

This paper is organized as follows. In Sec.~\ref{sec:EOB3PM} we take
full advantage of the most recent PM results appeared in the
literature~\cite{Cheung:2018wkq, Bern:2019nnu} and extend to 3PM order the
PM EOB Hamiltonian originally derived by Damour~\cite{Damour:2017zjx}
at 2PM order. In Sec.~\ref{sec:ener} we compare various binding-energy
curves obtained from PM, EOB and mixed PM-PN against each other and
NR, and discuss the implications of PM calculations for LIGO/Virgo
source modeling. Section~\ref{sec:concl} contains our final remarks 
and discusses future work.  In Appendix ~\ref{appendixA}, we first briefly discuss the
special role of the nonlocal-in-time (tail) part of the two-body
Hamiltonian at 4PN order. Then, we derive an extension of the 3PM EOB Hamiltonian computed in this paper 
to 4PN order, including such tail terms, using the 4PN EOB Hamiltonian in 
Ref.~\cite{Damour:2015isa}. In Appendix ~\ref{appendixB}, we start to explore 
how to improve the use of PM results in the EOB framework by presenting an 
alternative EOB Hamiltonian at 3PM order for circular orbits.

Henceforth, we work in units in which the speed of light $c=1$.

\section{An effective-one-body Hamiltonian at third post-Minkowskian order}
\label{sec:EOB3PM}

Damour \cite{Damour:2017zjx} and Cheung, Rothstein, and Solon (henceforth, CRS)
\cite{Cheung:2018wkq} have each given results for the
Hamiltonian governing the conservative dynamics of a
two-body system at 2PM order.  Damour's EOB Hamiltonian was deduced by
matching an ansatz to the gauge-invariant scattering angle function, 
first derived at 2PM order by Westpfahl \cite{Westpfahl:1985}, noting that
the complete local-in-time, gauge-invariant information content of the (conservative) 
Hamiltonian is encoded in the scattering angle computed from the Hamiltonian.
Westpfahl's 2PM result for the scattering angle has since been
rederived by Bjerrum-Bohr et al.\ \cite{Bjerrum-Bohr:2018xdl} by
applying the eikonal approximation to scattering amplitudes
for massive scalars exchanging gravitons at one-loop order.  The CRS
2PM Hamiltonian was deduced by directly matching between those same amplitudes and amplitudes computed
from an effective (classical) field theory.  As was noted later in Ref.~\cite{Bern:2019nnu}, and as we show in this section, the CRS 2PM Hamiltonian also leads to (and is determined by) Westpfahl's 2PM scattering angle.

Recently, Bern et al.\ \cite{Bern:2019nnu} (henceforth, BCRSSZ) have 
extended the computation of the classical-limit amplitudes to two-loop order, accomplished the matching to a 3PM Hamiltonian, and given the 3PM scattering angle.  Here we provide an independent derivation of the 3PM scattering angle from the 3PM 
Hamiltonian of Ref.~\cite{Bern:2019nnu}, and we use the scattering angle to extend the EOB Hamiltonian 
of Ref.~\cite{Damour:2017zjx} to 3PM order.  

We consider a two-body system composed of non-spinning black holes 
with rest masses $m_1$ and $m_2$, total mass $M = m_1+m_2$, 
reduced mass $\mu = m_1\,m_2/M$, and symmetric mass ratio $\nu = \mu/M$. 
The 3PM Hamiltonian of Ref.~\cite{Bern:2019nnu}, given
in the binary's center-of-mass frame and in an isotropic 
gauge, reads
\be
H(\bs r,\bs p)=H_0(\bs p^2)
+\sum_{n=1}^3\frac{G^n}{r^n}c_n(\bs p^2)
+\mc O(G^4),
\label{3PMH}
\ee
\be
H_0(\bs p^2)=\sqrt{m_1^2+\bs p^2}+\sqrt{m_2^2+\bs p^2},
\ee
where $\bs r$ and $\bs p$ are the radial separation and its conjugate momentum, respectively.  
 The functions $c_1$, $c_2$ and $c_3$ are explicitly given in Eqs.~(10) of 
BCRSSZ.  These functions determine (and are determined by) the coefficients in the 3PM scattering-angle function, 
as follows. (Henceforth, we refer to the Hamiltonian above as $H_{\rm 3PM}$.) 

Since we neglect black-hole's spins, the binary's orbital plane is fixed. We introduce polar coordinates $(r,\phi)$ in the orbital plane,
with conjugate momenta $(p_r,p_\phi\equiv L)$ satisfying the standard relation
\be\label{psq}
\bs p^2=p_r^2+\frac{L^2}{r^2}.
\ee
Note that $L=p_\phi$ is a constant of motion due to axial symmetry.  We denote with $E = H(\bs r,\bs p)=H(r,p_r,L)$ the total conserved energy of the binary system.  Using the Hamilton-Jacobi formalism, it can be shown 
(e.g., see Ref.~\cite{Damour:2017zjx}) that the total change in the angle coordinate $\phi$ for a scattering orbit is
given by
\be
\Delta\phi=\pi+\chi(E,L)=-2\int_{r_\mr{min}}^\infty dr\, \frac{\doe}{\doe L}p_r(r,E,L)\,,
\label{scattangle}
\ee
where $\chi$ is generally called the scattering angle, and vanishes for free motions. The radial momentum $p_r(r,E,L)$ is obtained by 
solving $H(r,p_r,L)=E$ for $p_r$ (taking the branch $p_r>0$ in Eq.~(\ref{scattangle})), while $r_\mr{min}$ is the appropriate 
root of $p_r=0$. 

The solution for $p_r$ resulting from the 3PM Hamiltonian (\ref{3PMH}) can be obtained from Eq.~(\ref{psq}) after we solve for $\bs p^2$, working perturbatively in $G$.  To conveniently express the dependence on the energy $E$, we define the quantities\footnote{ We notice that the true relative velocity at infinity for a scattering orbit is the $v$ in $\gamma=(1-v^2)^{-1/2}$, with $\gamma$ given in terms of the energy and masses by Eq.~\eqref{gG}.  The same quantity is called $\hat{\mc E}_\mr{eff}$ in Ref.~\cite{Damour:2016gwp}; at zeroth order (or, at infinity, to all orders), it is the quantity called $\sigma$ in BCRSSZ.  The $\Gamma$ in the right-hand side of Eq.~\eqref{gG} is called $h$ in Ref.~\cite{Damour:2017zjx}; at infinity, it is the variable $\gamma$ in BCRSSZ.}
\be
\gamma=\frac{E^2-m_1^2-m_2^2}{2m_1m_2},
\qquad
\Gamma\equiv\frac{E}{M}=\sqrt{1+2\nu(\gamma-1)}\,.
\label{gG}
\ee
From a straightforward calculation, using the results for $\{c_n(\bs p^2)\}_{n=1}^3$ from Eqs.~(10) of BCRSSZ, we find
\be
\label{psquared}
\bs p^2(r,E)=p_0^2(E)
+\sum_{n=1}^3\frac{G^n}{r^n}f_n(E)
+\mc O(G^4),
\ee
with
\begin{subequations}
\begin{align}
p_0^2&=\mu^2\frac{\gamma^2-1}{\Gamma^2},\\
f_1&=2\mu^2 M\frac{2\gamma^2-1}{\Gamma},\\
f_2&=\frac{3}{2}\mu^2M^2\frac{5\gamma^2-1}{\Gamma},\\
f_3&=\mu^2M^3\Bigg[\Gamma\frac{18\gamma^2-1}{2}
\nnm\\
&\quad-4\nu\gamma\frac{14\gamma^2+25}{3\Gamma}
+\frac{3}{2}\frac{\Gamma-1}{\gamma^2-1}(2\gamma^2-1)(5\gamma^2-1)
\nnm\\
&\quad-8\nu\frac{4\gamma^4-12\gamma^2-3}{\Gamma\sqrt{\gamma^2-1}}\sinh^{-1}\sqrt{\frac{\gamma-1}{2}}
\,\Bigg].
\end{align}
\end{subequations}
Combining Eqs.~\eqref{psq}, \eqref{scattangle}, and \eqref{psquared} and evaluating the integral, we find
\begin{alignat}{3}
\frac{\chi}{2}&=-\int_{r_\mr{min}}^\infty dr\frac{\doe}{\doe L}\sqrt{p_0^2-\frac{L^2}{r^2}+\sum_n\frac{G^n}{r^n}f_n}-\frac{\pi}{2}
\\\nnm
&=\frac{G}{L}\frac{f_1}{2p_0}+\frac{G^2}{L^2}\frac{\pi f_2}{4}
\\\nnm
&\quad+\frac{G^3}{L^3}\bigg(p_0f_3+\frac{f_1f_2}{2p_0}-\frac{f_1^3}{24p_0^3}\bigg)+\mc O(G^4).
\end{alignat}
Thus, the 3PM (half) scattering angle is given by
\be
\frac{1}{2}\chi(E,L)=\sum_{n=1}^3\bigg(\frac{GM\mu}{L}\bigg)^n\chi_n(E)+\mc O(G^4),
\ee
with coefficients
\begin{subequations}\label{chi_n}
\begin{align}
\chi_1&=\frac{2\gamma^2-1}{\sqrt{\gamma^2-1}},\\
\chi_2&=\frac{3\pi}{8}\frac{5\gamma^2-1}{\Gamma},\\
\chi_3&=\frac{64\gamma^6-120\gamma^4+60\gamma^2-5}{3(\gamma^2-1)^{3/2}}\nnm
\\
&\quad-\frac{4}{3}\frac{\nu}{\Gamma^2}\gamma\sqrt{\gamma^2-1}(14\gamma^2+25)
\nnm\\
&\quad-8\frac{\nu}{\Gamma^2}(4\gamma^4-12\gamma^2-3)\sinh^{-1}\sqrt{\frac{\gamma-1}{2}},
\end{align}
\end{subequations}
which agrees with Eq.~(12) of BCRSSZ.
When we take the limit $\nu\to0$ at fixed $\gamma$, implying $\Gamma\to1$, the scattering angle reduces to the one for a test particle with energy-per-mass $\gamma$ and angular-momentum-per-mass $L/\mu$, following a geodesic in a Schwarzschild spacetime with mass $M$.
Note that $\chi_1$ is the same as the Schwarzschild value; $\chi_2$ is the Schwarzschild value over $\Gamma$. 
The first line of $\chi_3$ coincides with its Schwarzschild value. 

Damour has shown in Ref.~\cite{Damour:2017zjx} that an EOB Hamiltonian valid at 3PM order (for the conservative dynamics) can be obtained directly from the scattering-angle coefficients. The real EOB Hamiltonian $H^\mr{EOB}(\bs r,\bs p)$ is given in terms of the effective Hamiltonian $H^\mr{eff}(\bs r,\bs p)$ via the EOB energy map~\cite{Buonanno:1998gg},
\begin{equation}
H^{\rm EOB} = M\,\sqrt{ 1 + 2 \nu\,\left (\frac{H^{\rm eff}}{\mu} - 1 \right ) }\,,
\label{3PMEOBH}
\end{equation}
and $H^\mr{eff}$ reduces to the Hamiltonian for Schwarzschild geodesics $H_\mr S$ as $\nu\to0$.
The Schwarzschild-geodesic Hamiltonian (for a test particle of mass $\mu$) is given in Schwarzschild coordinates in the equatorial plane, with $H_\mr S(\bs r,\bs p) \equiv H_\mr S(r,p_r,L)$, by
\be
H_\mr S^2 =
\left(1-\frac{2GM}{r}\right)\,\left [\mu^2+\frac{L^2}{r^2}+\left(1-\frac{2GM}{r}\right)p_r^2\right ]\,.
\label{Hs}
\ee
Defining the reduced (dimensionless) quantities
\begin{gather}\label{reducedHu}
  \hat H^\mr{eff}=\frac{H^\mr{eff}}{\mu},\quad \hat H_\mr S=\frac{H_\mr S}{\mu},\quad u=\frac{GM}{r},\\
  \hat{p}_r = \frac{p_r}{\mu},\quad l \equiv \hat{p}_\phi = \frac{L}{GM\mu},
\end{gather}
the effective Hamiltonian of Ref.~\cite{Damour:2017zjx}---which we will refer to as the post-Schwarzschild (PS) effective Hamiltonian---is given through 3PM order by Eq.~(5.13) of
Ref.~\cite{Damour:2017zjx} as
\begin{align}
(\hat H^\mr{eff,PS})^2&=
\hat H_\mr S^2
+(1-2u)\Big[u^2q_\mr{2PM}+u^3q_\mr{3PM}
 +\mc O(G^4)\Big],
\label{3PMeffH}\\
\hat H_\mr S^2 &= (1-2u) \left[1 + l^2 u^2 + (1-2u) \hat{p}_r^2 \right], \label{HShat}
\end{align}
where the functions $q_\mr{2PM}(\hat H_\mr S,\nu)$ and $q_\mr{3PM}(\hat H_\mr S,\nu)$ are determined by the scattering-angle coefficients via Eqs.~(5.6) and (5.8) of Ref.~\cite{Damour:2017zjx}. (Notice the absence of a $q_\mr{1PM}(\hat H_\mr S,\nu)$ term, which vanishes identically in the EOB formulation. Indeed, the energy map \eqref{3PMEOBH} applied to the unmodified Schwarzschild-geodesic Hamiltonian \eqref{Hs} precisely reproduces the two-body dynamics at 1PM order \cite{Damour:2016gwp,Damour:2017zjx}.)

Inserting our coefficients (\ref{chi_n}) into those equations yields
\begin{subequations}\label{q2q3}
\begin{align}
q_\mr{2PM}&=\frac{3}{2}(5\hat H_\mr S^2-1)\Bigg(1-\frac{1}{\sqrt{1+2\nu(\hat H_\mr S-1)}}\Bigg),\label{qtwo}
\\
q_\mr{3PM}&=-\frac{2\hat H_\mr S^2-1}{\hat H_\mr S^2-1}q_\mr{2PM}
+\frac{4}{3}\nu\hat H_\mr S\frac{14\hat H_\mr S^2+25}{1+2\nu(\hat H_\mr S-1)}\nnm\\
&\quad+\frac{8\nu}{\sqrt{\hat H_\mr S^2-1}}\frac{4\hat H_\mr S^4-12\hat H_\mr S^2-3}{1+2\nu(\hat H_\mr S-1)}\sinh^{-1}\sqrt{\frac{\hat H_\mr S-1}{2}}. \label{qthree}
\end{align}
\end{subequations}
The resultant 3PM EOB Hamiltonian, $H^{\rm EOB}_{\rm 3PM}$, for the two-body description, is obtained by plugging (\ref{3PMeffH}) into Eq.~(\ref{3PMEOBH}). The $H^{\rm EOB}_{\rm 3PM}$ and BCRSSZ Hamiltonians are
equivalent in the sense that they lead to the same scattering angle when
expanded at 3PM order.  In the PN expansion, they are both complete up to 2PN order.  We discuss in Appendix \ref{appendixA} how to augment the above 3PM EOB Hamiltonian with additional PN information at 3PN and 4PN orders.

\begin{figure*}
	\includegraphics[width=\linewidth]{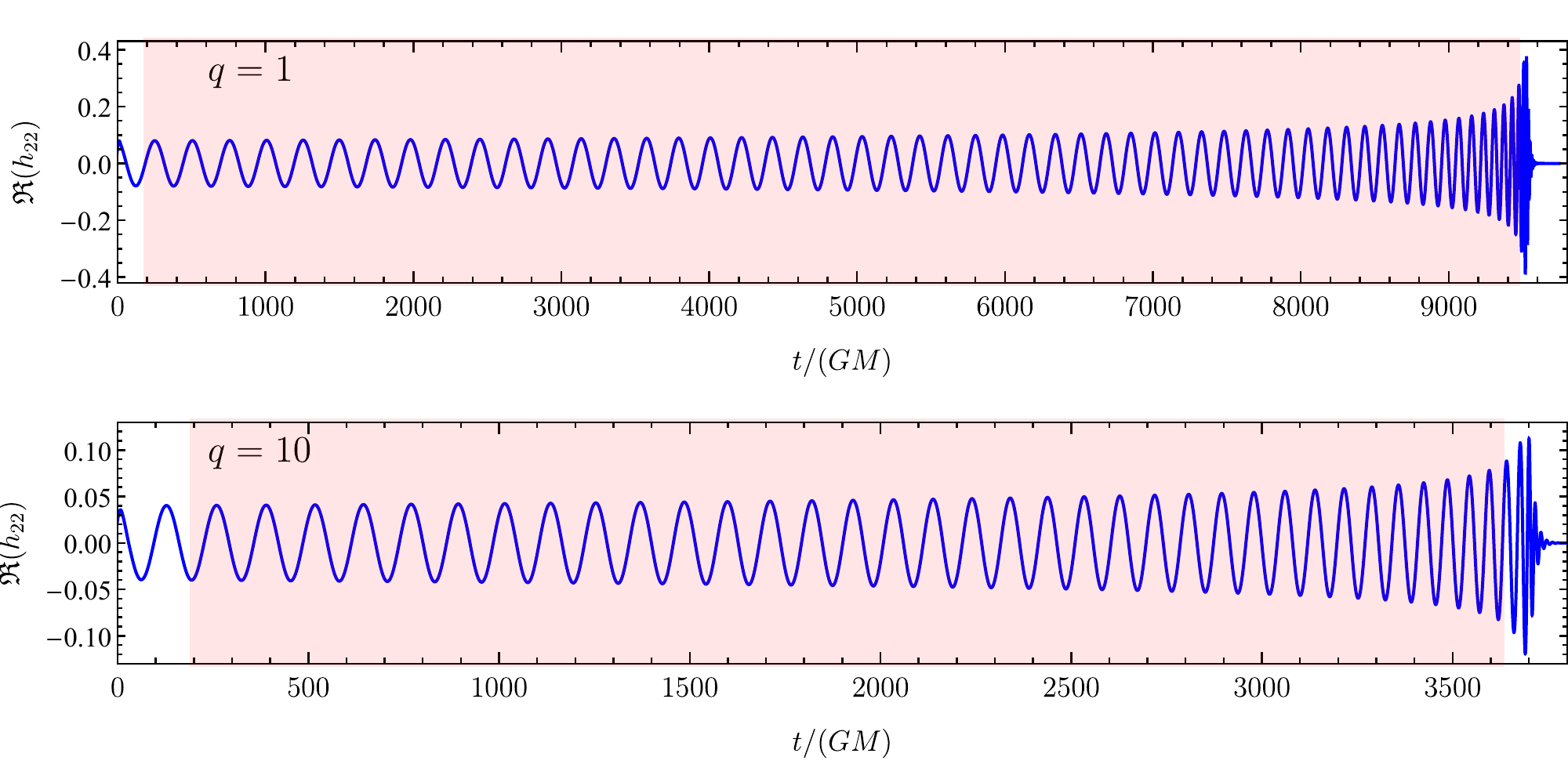}
	\caption{{\bf NR simulations.} In this paper the energetics of various approximants are compared against two NR simulations of non-spinning binary black holes produced by the Simulating eXtreme Spacetimes (SXS) collaboration~\cite{Mroue:2013xna,Chu:2015kft}. The top (bottom) panel shows the waveform (more specifically the real part of the $l=m=2$ mode of the strain, $\mathfrak{R}(h_{22})$) of the simulation with mass-ratio $q=1$ ($q=10$), identified in the SXS catalog as SXS ID: 0180 (SXS ID: 0303). In both panels, the red shading shows the 
segment of the simulation used for the binding-energy's comparisons in all figures of this paper.
		\label{fig:waveforms}}
\end{figure*}

\section{Energetics of binary systems with post-Minkowskian Hamiltonians}
\label{sec:ener}

Gravitational waveforms emitted by inspiraling binaries are constructed from the binary's binding energy and GW energy flux (e.g., see Refs.~\cite{Buonanno:2009zt}). To assess the relevance for LIGO/Virgo analysis of the recently derived conservative two-body dynamics in PM theory, we compute one of these building blocks, the binding energy, for a variety of PM, PN and EOB approximants. We then compare these with results from NR simulations.

\begin{figure*}[t]
	\includegraphics[width=\columnwidth]{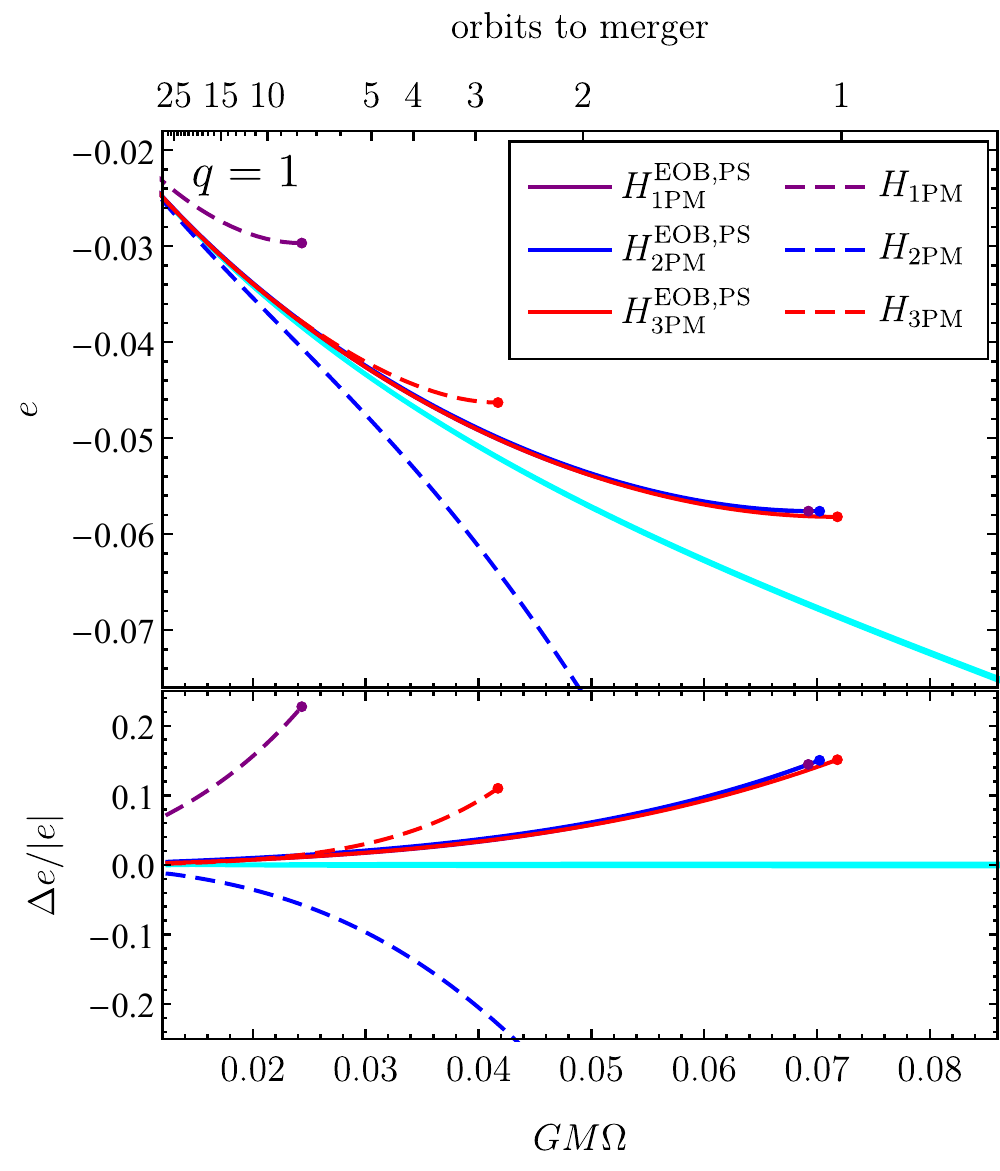}
	\hspace{12pt}
	\includegraphics[width=\columnwidth]{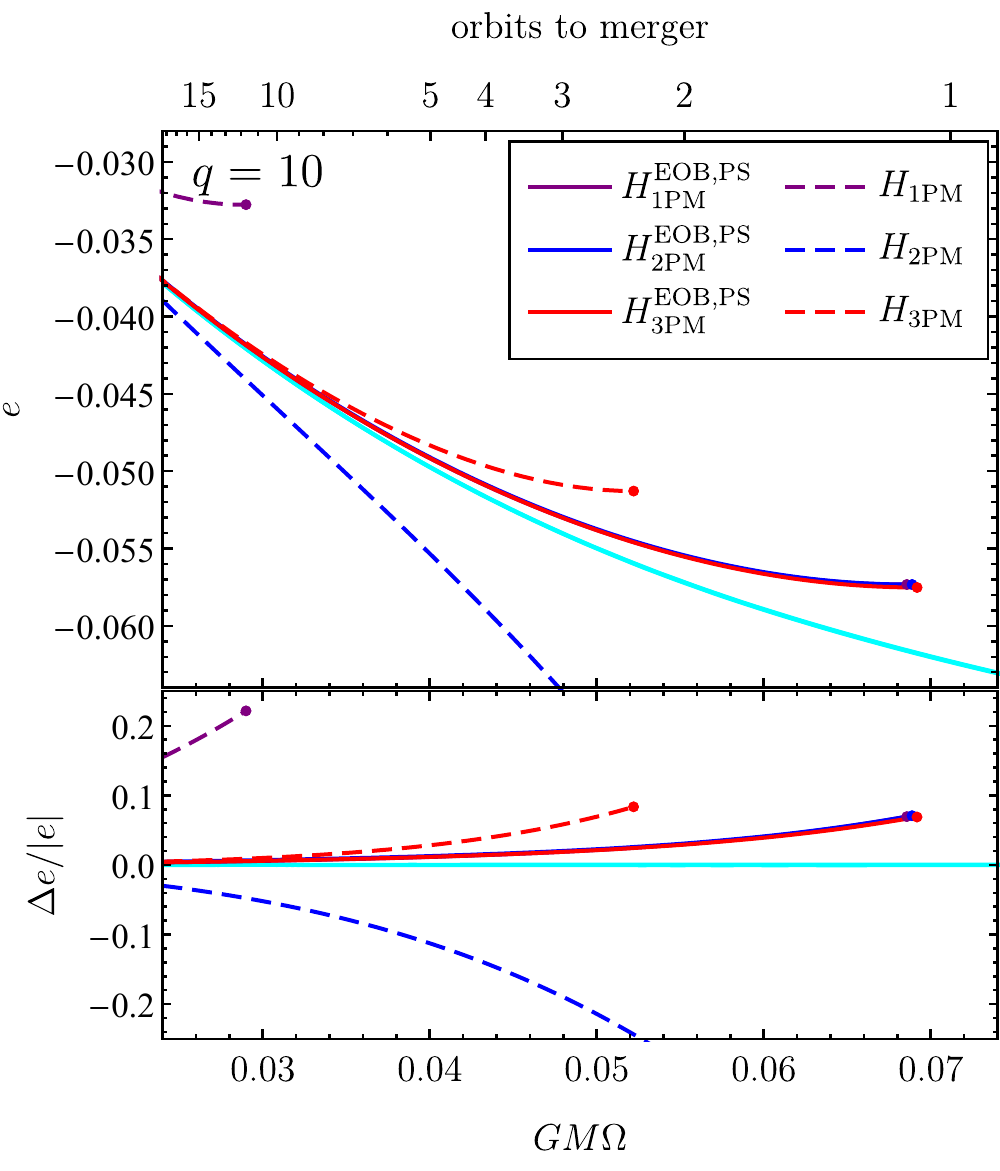}
	\caption{{\bf Energetics of PM Hamiltonians.} We compare to NR the binding energy as a function of orbital frequency $GM\Omega$ from both PM and PM-EOB Hamiltonians for a nonspinning binary black hole with mass ratio $q=1$ (left panel) and $q=10$ (right panel). The dots at the end of the curves mark the ISCOs, when present in the corresponding two-body dynamics. The NR binding energy and its error are in cyan. The top $x$-axis shows the number of orbits until merger. In the lower panel we show the fractional difference between the approximants and the NR result. 
		\label{fig:energywPM}}
\end{figure*}

\begin{figure*}[t]
	\includegraphics[width=\columnwidth]{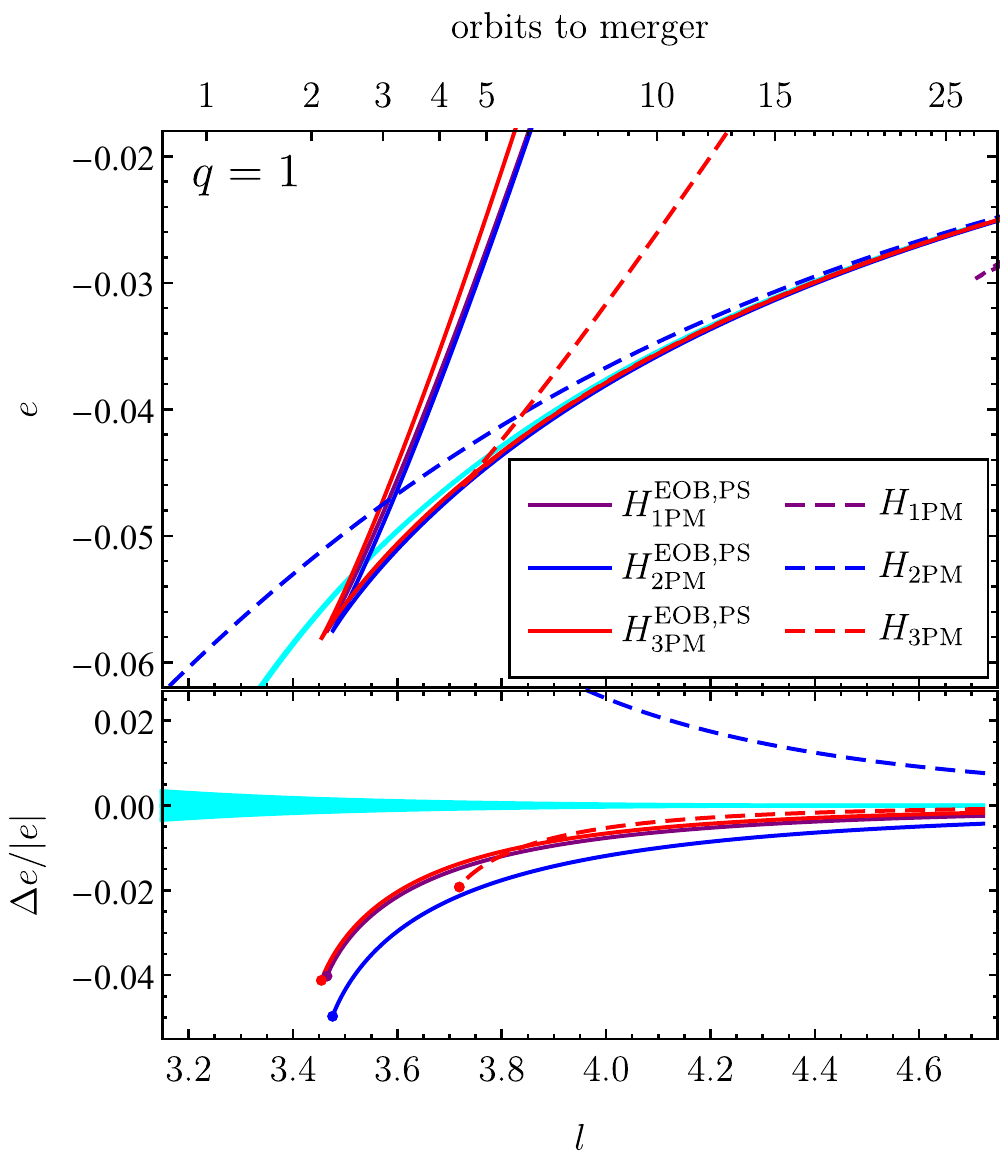}
	\hspace{12pt}
	\includegraphics[width=\columnwidth]{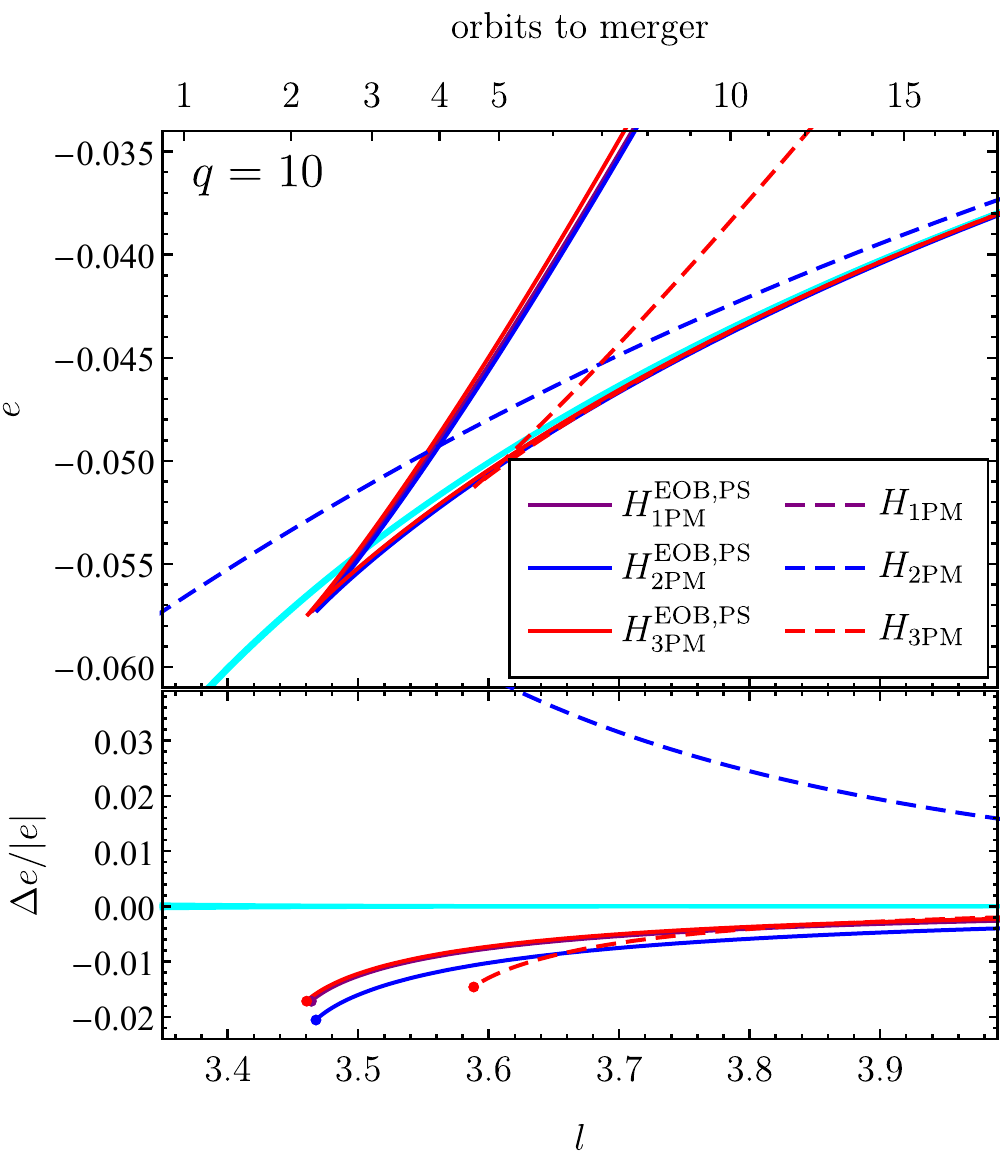}
	\caption{{\bf Energetics of PM Hamiltonians.} Same as in Fig.~\ref{fig:energywPM} but versus the dimensionless angular momentum $l=L/(G \mu M)$. The cusps signal the presence of the ISCO, where the branches of stable and unstable circular-orbit solutions meet. Note that the orbital-frequency range in the plots ends about $1.4$ and $1.8$ GW cycles, for $q=1$ and $q=10$, respectively, before the two black holes merge. \label{fig:energyJPM}}
\end{figure*}

We recall that the total energy $E$, linear momentum $\vct{P}$, and angular momentum $\vct{L}$ of a gravitating two-body system in an asymptotically flat spacetime are nearly\footnote{The quantities $E$, $\vct{P}$, and $\vct{L}$ are only invariant up to the fixing of a frame at infinity.} gauge-invariant quantities. It is convenient to introduce the dimensionless binding energy $e \equiv (E - M) / \mu$ and angular momentum $l \equiv |\vct{L}| / (G M \mu)$.  For an inspiraling binary of non-spinning black holes, the energy and angular momentum monotonically decrease over time and trace out a curve $e(l)$ for each set of binary parameters. This $e(l)$-curve is a gauge-invariant relation that can be used to compare analytic predictions in PN, PM and EOB frameworks against numerical-relativity (NR) results~\cite{Damour:2011fu,LeTiec:2011dp,Nagar:2015xqa,Ossokine:2017dge}. In absence of radiation reaction and for circular orbits, the relation $e(l)$ encodes the conservative dynamics.  For quasi-circular inspirals, the $e(l)$-relation still depends most sensitively on the conservative dynamics.  Hence, it is a good indicator for the accuracy of the binding energy derived from the PN, PM and EOB Hamiltonians for circular orbits.

Now, $E$ and $\vct{L}$ are not directly extracted from NR simulations as a function of time, but instead it is the gravitational radiation leaving the binary system (more precisely, the ``news function'') that is extracted~\cite{Damour:2011fu,Ossokine:2017dge}. The radiated energy and angular momentum fluxes as functions of time can then be integrated to yield the energy $E$ and angular momentum $\vct{L}$ of the binary at a given (retarded) time, which can then be combined into the relation $e(l)$~\cite{Damour:2011fu,Ossokine:2017dge}. The integration constants can be adjusted to match $E$ and $\vct{L}$ at the initial time of the simulation~\cite{Damour:2011fu,Nagar:2015xqa}, which has the disadvantage that one has to accurately track the fluxes during the initial junk-radiation phase.  A better approach is to fix the integration constants to match the energy and angular momentum of the final (merged) black hole; but even in this case, a further tuning of the integration constants is needed to achieve agreement with analytical models of the early inspiral (e.g., see Ref.~\cite{Ossokine:2017dge}). In the following, we use the binding energy from NR simulations as extracted and tuned in Ref.~\cite{Ossokine:2017dge}.

Similarly, the relation $e(l)$ can be obtained by solving the Hamilton equations with radiation-reaction effects for a given PN, PM and EOB Hamiltonian (e.g., as done in Refs.~\cite{Damour:2011fu,LeTiec:2011dp}).  However, for most of the analysis in this section, we neglect radiation-reaction effects (which have been shown to make EOB Hamiltonians accurate past the last-stable orbit and all the way down to merger~\cite{Damour:2013tla}), and construct the $e(l)$ curves using an adiabatic sequence of circular orbits instead. For this reason, we should not expect exact agreement with the NR results, which do include radiation-reaction effects. Our motivation for this choice is that $e(l)$ only depends on the Hamiltonian model (and not the radiation-reaction model), so it is easier to interpret our results and put them into context for future improvements of the Hamiltonian (e.g., when higher-order PN and PM results become available). More importantly, previous investigations~\cite{Damour:2011fu,LeTiec:2011dp} have shown that at least until the innermost-stable circular orbit (ISCO, where we terminate the comparison with NR results), the difference between the binding energy computed from a sequence of circular orbits and from a quasi-circular inspiral is not very large (typically no more than 5-10\%, as we discuss below and in Fig.~\ref{fig:evolq1}).

%\begin{table}
%	\caption{{\bf NR simulations.} We summarize the NR simulations used here and produced by the Simulating %eXtreme Spacetimes (SXS) collaboration~\cite{Mroue:2013xna,Chu:2015kft}. We list the SXS ID of the simulations, the mass-ratio $q$, the angular momentum at merger $l_{\rm merger}$, the orbital frequency at merger $(GM\Omega)_{\rm merger}$. Then, defining the ranges $[(GM\Omega)_{\rm start},(GM\Omega)_{\rm end}]$ and $[(l_{\rm start},l_{\rm end})]$ of orbital frequency  and angular momentum displayed in Figs.~\ref{fig:energywPM}--\ref{fig:vanPMvsPNq1}, we also list the number of orbital cycles 
%$N_{\rm orb}$ from $(GM\Omega)_{\rm start}$ to $(GM\Omega)_{\rm end}$, and from $(GM\Omega)_{\rm start}$ to the binary black-hole merger.
%		\label{table:simulations}
%	}
%	\begin{ruledtabular}
%		\begin{tabular}{rcc}
%			SXS ID						&	0180			&	0303 	\\
%			\hline
%			q							&	1			%	&	10		\\
%			$l_{\rm merger}$			&	2.802			&	3.209	%\\
%			$(GM\Omega)_{\rm merger}$	&	0.180			&	0.150	\\
%			$[(GM\Omega)_{\rm start},(GM\Omega)_{\rm end}]$ 	&	(0.012, 0.086)	&	(0.024, 0.074)	\\
%			$(l_{\rm start},l_{\rm end})$			&	(3.400, 4.750)	&	(3.400, %3.990)	\\
%			$N_{\rm orb}$ from start to end		&	27.1			&	17.5	%	\\
%			$N_{\rm orb}$ from start to merger	&	27.8			&	18.4	%	\\
%		\end{tabular}
%	\end{ruledtabular}
%\end{table}

\begin{table}
	\caption{{\bf Two-body Hamiltonians.} A summary of the Hamiltonians used in this paper to compute the binding energy and compare it with NR predictions. 
		\label{table:models}}
	\begin{ruledtabular}
		\begin{tabular}{lp{5cm}p{1.5cm}}
			$H_{m\rm PM}$
				&	PM Hamiltonian	& \cite{Cheung:2018wkq,Bern:2019nnu} \\
			$H_{m\rm PM}^{\rm EOB,PS}$					
				&	PM EOB Hamiltonian & \cite{Damour:2017zjx} and this paper \\
			$H^{\rm EOB,PS}_{m{\rm PM} + n{\rm PN}}$
				&       PM EOB Hamiltonian with PN information when $n\geq m$ & \cite{Damour:2017zjx} and this paper	\\
			$H_{n\rm PN}^{\rm EOB}$	
				&	PN EOB Hamiltonian used in LIGO/Virgo data-analysis & \cite{Buonanno:1998gg,Damour:2000we,Damour:2015isa} \\
 		        $H_{3\rm PM}^{\rm EOB,\widetilde{PS}}$ & alternative 3PM EOB Hamiltonian & this paper\\
                        $H_{n\rm PN}$	& PN Hamiltonian & \cite{Jaranowski:1997ky}	\\
		\end{tabular}
	\end{ruledtabular}
\end{table}

In the absence of radiation reaction, the Hamilton equations for a generic Hamiltonian $H(r, p_r, L)$ describing a two-body system of nonspinning black holes read,
\begin{subequations}
	\begin{align}
	\dot{r} &= \frac{\partial H}{\partial p_r}\,, &
	\dot{L} &= - \frac{\partial H}{\partial \phi}=0\,, \\
	\Omega \equiv \dot{\phi} &= \frac{\partial H}{\partial L}\,, &
	\dot{p}_r &= - \frac{\partial H}{\partial r} \,.
	\end{align}
\end{subequations}
Note that $L \equiv G M \mu l = \text{const}$. For circular orbits, $p_r = 0$, $r = r_\text{circ} = \text{const}$. Furthermore, $\dot{p}_r=0$ and consequently it follows from the Hamilton equations that $({\partial H}/{\partial r})_{r = r_\text{circ}}=0$, which determines $r_\text{circ}(l)$ and hence the circular-orbit relation $e(l) \equiv  [H(r_\text{circ}(l), 0,  l) - M]/\mu$. The relation $\Omega(l) \equiv (\partial H/\partial L)_{r = r_\text{circ}} = e'(l)$ determines a second gauge-invariant relation. Inverting this relation gives $l(\Omega)$, which can be combined with $e(l)$ to give $e(\Omega)$.

\begin{figure*}[t]
	\includegraphics[width=\columnwidth]{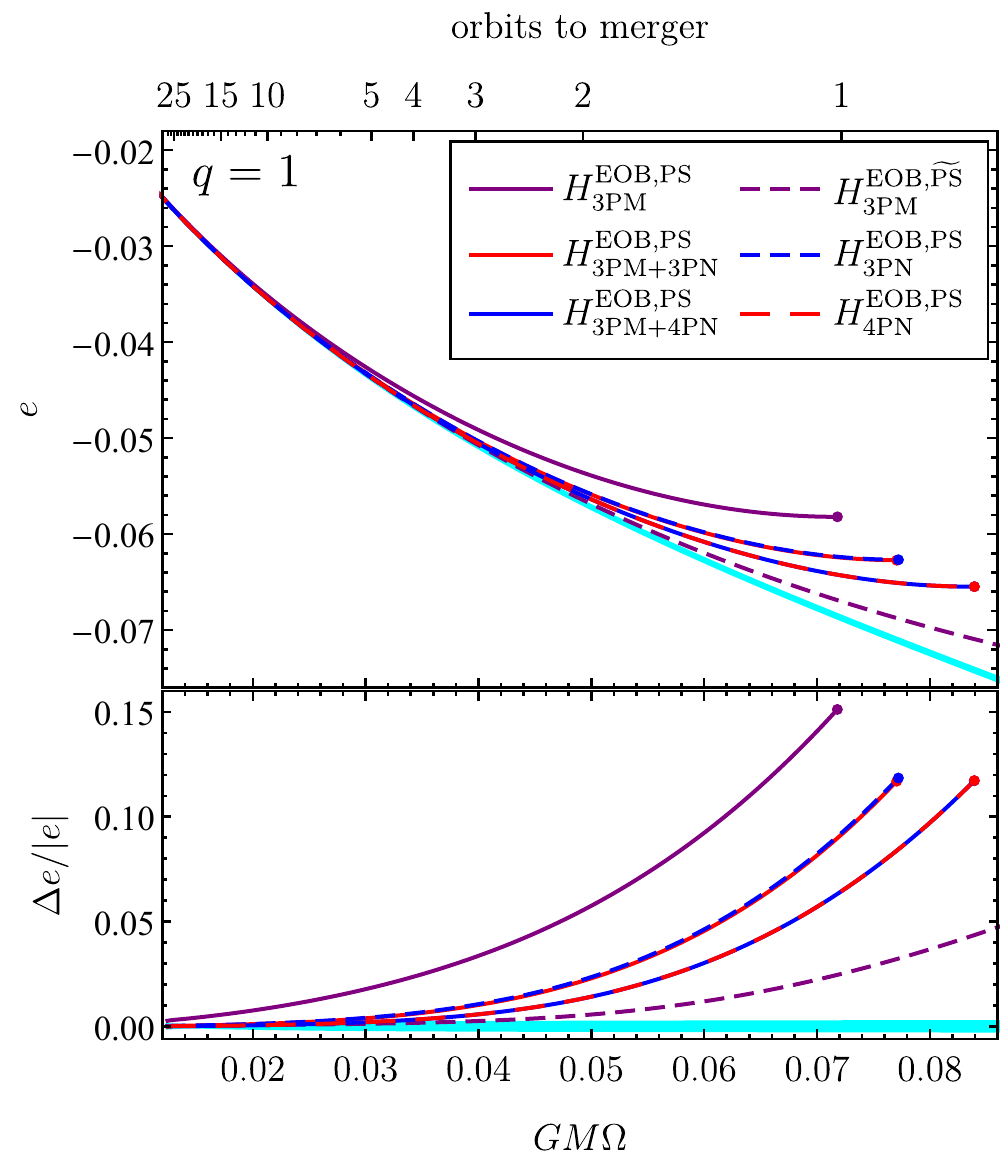}
	\hspace{12pt}
	\includegraphics[width=\columnwidth]{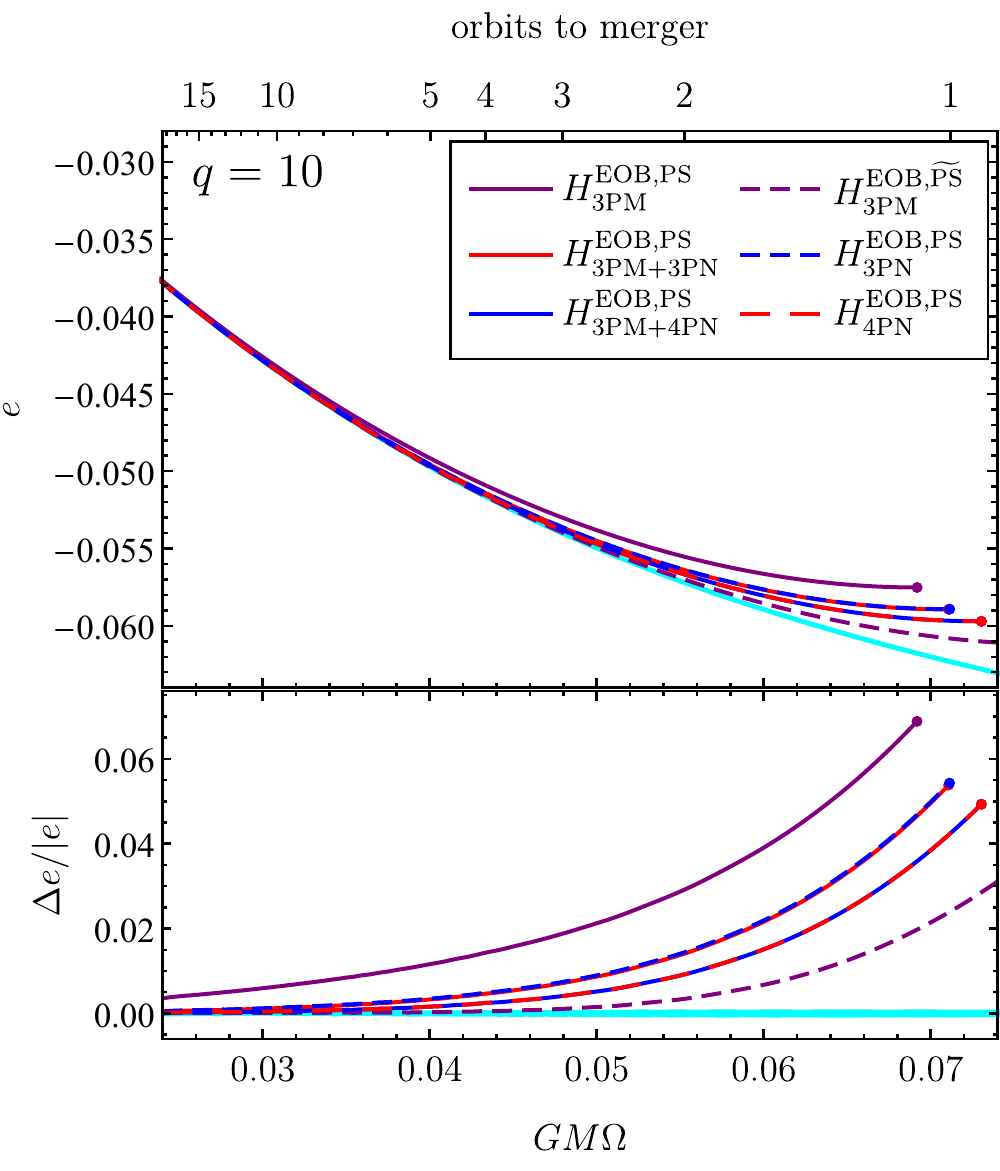}
	\caption{{\bf Energetics of PM Hamiltonians augmented by PN information.} Same as in Fig.~\ref{fig:energywPM} but now we compare to NR the binding energy of PM EOB Hamiltonians augmented by PN information. Notice that adding 3PM information at 3PN or above does not lead to a visible difference from plain PN EOB Hamiltonians (the 3PM-3PN and 3PN curves, as well as the 3PM-4PN and 4PN ones, are essentially on top of each other). Also included is a curve for an alternative 3PM EOB Hamiltonian,  $H^{\rm EOB,\widetilde{PS}}_{3 \rm PM}$, 
derived in Appendix \ref{appendixB}.
          \label{fig:energymixed}}
\end{figure*}

Given a Hamiltonian there are different ways to determine the $e(l)$
and $e(\Omega)$ relationships. For example, we can solve for them
order-by-order in a systematic PN expansion yielding $e$ and $l$ as a
power series in $(GM\Omega)^{2/3}$ (e.g., see Eq.~(232) in Ref.~\cite{Blanchet:2013haa}). 
However, we note that if we were just expanding the binding energy computed from the 
PM Hamiltonian in powers of $(GM\Omega)^{2/3}$ and then truncating it, all extra information obtained
through the PM approximation would be lost. Nevertheless, performing 
such PN expansion of the binding energy provides an important consistency
check between different Hamiltonians --- for example one readily verifies that
starting from the 3PM Hamiltonian of Ref.~\cite{Bern:2019nnu} one
obtains the well known PN circular orbit relations $e(\Omega)$ and
$l(\Omega)$ to order  $(GM\Omega)^{2}$. To gauge the
additional information present in the PM Hamiltonians we opt for a
different approach, where we treat the various PN, EOB and PM 
approximants as exact Hamiltonians and determine the relations $e(l)$
and $e(\Omega)$ numerically (i.e. without any further expansions). 
We discuss in Fig.~\ref{fig:vanPMvsPNq1} below, differences in the PN 
binding energy when we build it from the exact (unexpanded)  PN Hamiltonian and the 
systematically PN expanded one.  
   
We consider only stable (and marginally stable) circular orbits, for which the Hamiltonian is minimal, $ 0 \leq ({\partial^2 H}/{\partial r^2})_{r = r_\text{circ}}$. Here equality corresponds to a saddle point of the Hamiltonian, which indeed exists for most --- but not all --- of the Hamiltonians under investigation. This is the well-known ISCO and corresponds to an angular momentum $l = l_\text{ISCO}$. 

For simplicity, we restrict the discussion to NR simulations of nonspinning binary black holes with 
mass ratios $q = 1$ and $10$~\cite{Ossokine:2017dge}. In Fig.~\ref{fig:waveforms} we display the 
NR waveforms. Those simulations span about $56$ and $36$ GW cycles (corresponding to $\sim 28$ and $\sim 18$ orbital cycles), for $q=1$ and $q=10$, respectively, before merger. 
We highlight in Fig.~\ref{fig:waveforms} the portion of the waveform that we use to compare with 
the binding-energy approximants. As can be seen, the comparisons with NR extend up to 
about $1.4$ and $1.8$ GW cycles, for $q=1$ and $q=10$, respectively, before the two black holes merge. 
Thus, our comparisons of analytic models to NR predictions extend to the late inspiral of a binary evolution, a stage characterized by high velocity and strong gravity.

We compare NR predictions against analytic results obtained with PM, EOB and PN Hamiltonians, summarized in Table~\ref{table:models}. Notably, we compute results with the Hamiltonian at $m {\rm PM}$ orders with $m = 1, 2, 3$~\cite{Cheung:2018wkq, Bern:2019nnu} (labeled $H_{m\rm PM}$),  and with the EOB Hamiltonian of Refs.~\cite{Damour:2016gwp,Damour:2017zjx} and this paper at $m {\rm PM}$ orders with $m = 1, 2,3$ (labeled $H^{\rm EOB, PS}_{m\rm PM}$). We also compare results with the PM EOB Hamiltonian augmented with PN results up to 4PN order (labeled $H^{\rm EOB,PS}_{m{\rm PM} + n{\rm PN}}$), as derived in Appendix \ref{appendixA}. Furthermore, the (original) EOB Hamiltonian employed in LIGO/Virgo data 
analysis~\cite{Taracchini:2013rva,Bohe:2016gbl} is built from the EOB Hamiltonian of Refs.~\cite{Buonanno:1998gg,Damour:2000we,Damour:2015isa}, and it resums perturbative PN results differently from the PM EOB Hamiltonian. To understand the impact of the different 
resummation, and also highlight the accuracy that PM results would need to achieve in order to motivate their use in waveform modeling, we also show results with such an EOB Hamiltonian (labeled $H^{\rm EOB}_{n{\rm PN}}$). Finally, we also employ the PN Hamiltonian 
from Ref.~\cite{Jaranowski:1997ky} (labeled $H_{n\rm PN}$), and an alternative 3PM EOB Hamiltonian presented for circular orbits 
in Appendix \ref{appendixB} (labeled $H^{\rm EOB,\widetilde{PS}}_{\rm 3PM}$).

\begin{figure}
	\includegraphics[width=\columnwidth]{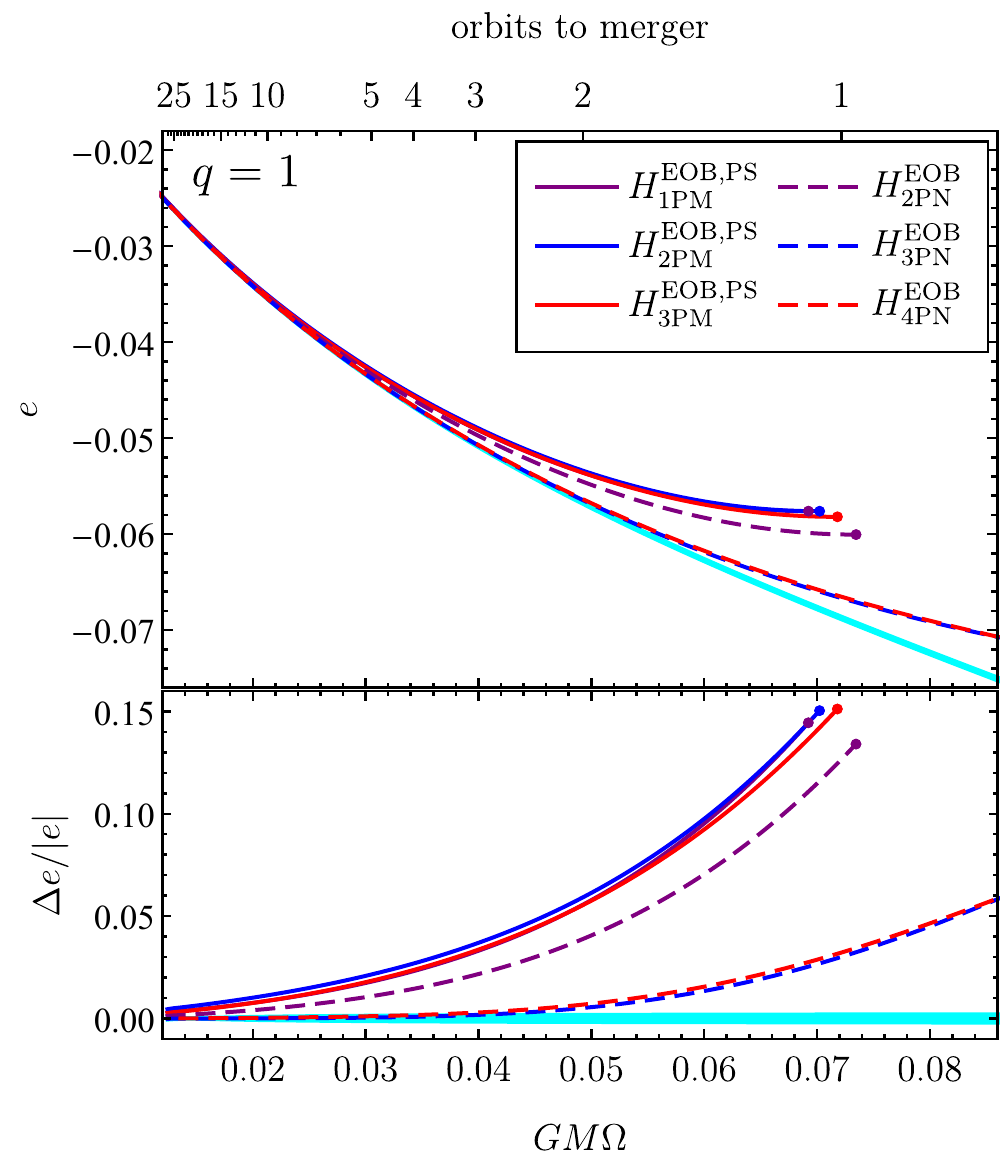}
	\caption{{\bf Energetics of PM EOB Hamiltonian and the EOB Hamiltonian used in LIGO/Virgo data-analysis.} Same as in Fig.~\ref{fig:energywPM} and Fig.~\ref{fig:energymixed}, but now we show how the $H^{\rm EOB,PS}_{m \rm PM}$ Hamiltonian compares with the (original)  $H^{\rm EOB}_{n \rm PN}$ Hamiltonian currently employed at 4PN order to 
			build waveform models for LIGO/Virgo data-analysis. We observe that $H^{\rm EOB}_{n \rm PN}$ Hamiltonians still produce $e(\Omega)$-curves substantially closer to NR result than the 3PM approximant.
		\label{fig:PMvsclassicEOB}}
\end{figure}

\begin{figure}
	\includegraphics[width=\columnwidth]{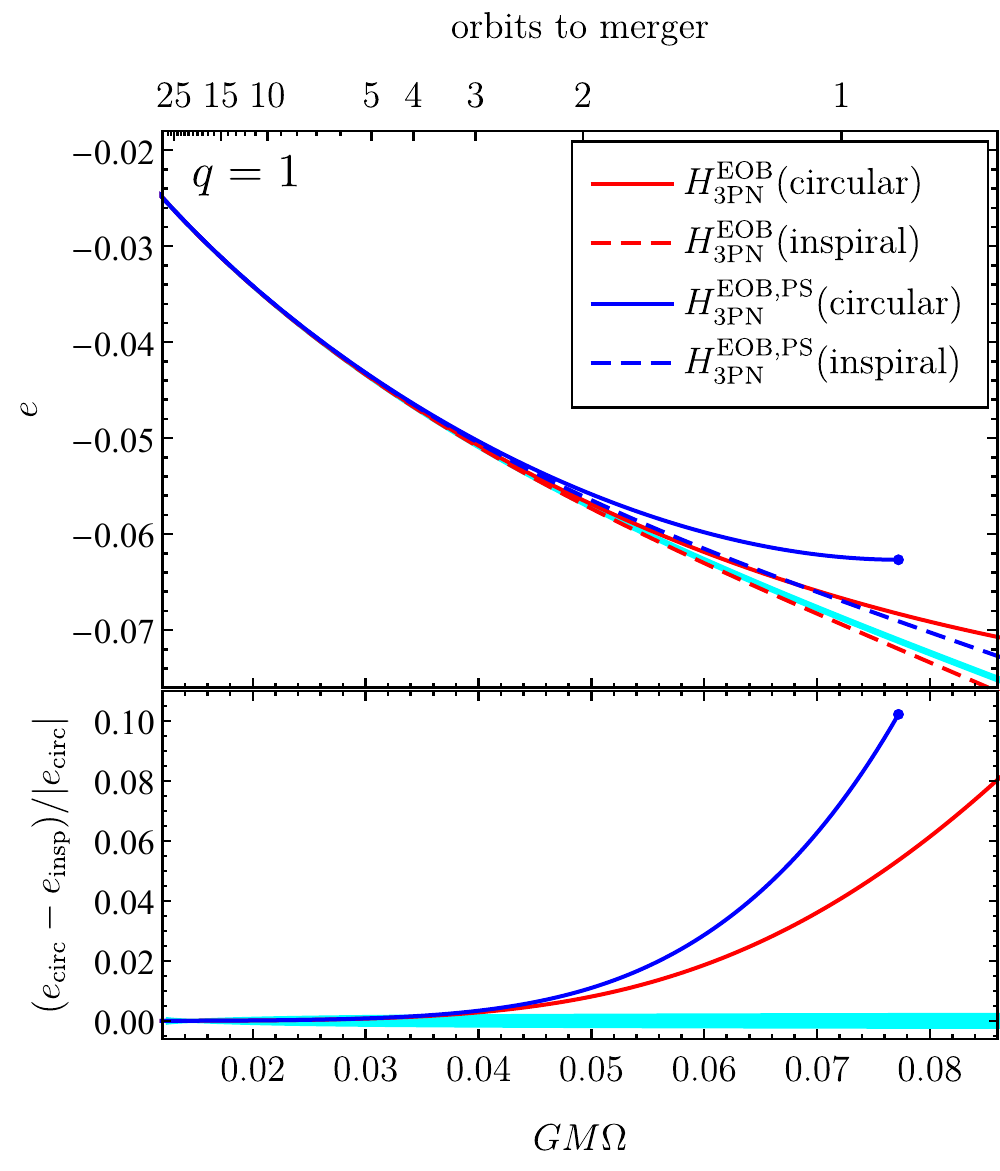}
	\caption{{\bf Energetics of circular versus inspiral  PN approximants.}
		We show the binding energy obtained from the $H^{\rm EOB}_{3 \rm PN}$ and $H^{\rm EOB,PS}_{n \rm PN}$ either through an adiabatic sequence of circular orbits or numerically evolving the Hamilton equations with a suitable radiation-reaction 
			force for a quasi-circular inspiral.  The bottom panel shows the relative difference between circular and inspiral curves. This gives an indication of the magnitude of the impact which should be kept in mind when interpreting the other figures. By comparison the size of the NR error --- indicated in gray --- is very small.
		\label{fig:evolq1}}
\end{figure}

In Figs.~\ref{fig:energywPM} and \ref{fig:energyJPM} we compare the
binding energy computed in NR with the ones from PM and PM EOB
Hamiltonians versus either the binary's orbital frequency (Fig.~\ref{fig:energywPM}) or angular momentum
(Fig.~\ref{fig:energyJPM}), for mass ratios $q=1$ and $q=10$. We
clearly see the improvement of the PM binding energy from 1PM to 3PM,
especially at low frequency. The PM-EOB binding energies
  generally show better agreement with NR, but they have a much
smaller range of variation from 1PM to 3PM. The 3PM result does
  slightly better than 1PM, while 2PM is worse than the other two.
Overall those results demonstrate the value and relevance of pushing
PM calculations at higher order, and of further exploring how to
  use PM results to improve EOB models.

  To understand the impact of PM calculations for LIGO/Virgo analyses,
  it is important to compare the PM binding energy with current
  approximants used to build waveform models. 
  Let us emphasize again
  that perturbative PM calculations (weak-field/fast-motion), suitable
  for unbound/scattering orbits, are not necessarily expected to
  improve, when available at low PM orders, the predictions obtained
  in perturbative PN calculations (weak-field/slow-motion), suitable
  for bound/inspiraling orbits, which are the LIGO/Virgo GW
  sources. It is
  instructive to understand how the PM binding
  energy compares with the PN binding energy, which at $n$PN order we
  expect to be more accurate than the one at $n$PM order. For this
  study we restrict to the 3PM EOB Hamiltonian and augment it with 3PN
  and 4PN information, as derived explicitly in
  Appendix~\ref{appendixA}. We display results in
  Fig.~\ref{fig:energymixed}. Interestingly, the figure shows that the
  mixed PM-PN Hamiltonian does not improve much over a PN Hamiltonian.
  This means that currently the known PM Hamiltonian does not improve
  in accuracy compared to PN ones (as usual, regarding NR as the
  ``true'' result). However, it is important to note that so far the PM
  information has been incorporated into the EOB Hamiltonian in one
  particular way, as proposed in Ref.~\cite{Damour:2017zjx}, in the
  $H^{\rm EOB,PS}_{m \rm PM}$ curves.  We note in
  Fig.~\ref{fig:energymixed} that one alternatively resummed EOB-3PM
  Hamiltonian, $H^{\rm EOB,\widetilde{PS}}_{3 \rm PM}$, defined in
  Appendix~\ref{appendixB}, shows better agreement with NR.  In the
  PM expansion, this Hamiltonian is perturbatively equivalent to
  $H^{\rm EOB,{PS}}_{3 \rm PM}$ up to 3PM order, i.e., they differ
  only by 4PM-order terms.  The variation between those two curves
  thus gives some indication of the variability expected from 4PM
  order, and motivates calculations at higher PM order. 

\begin{figure}
	\includegraphics[width=\columnwidth]{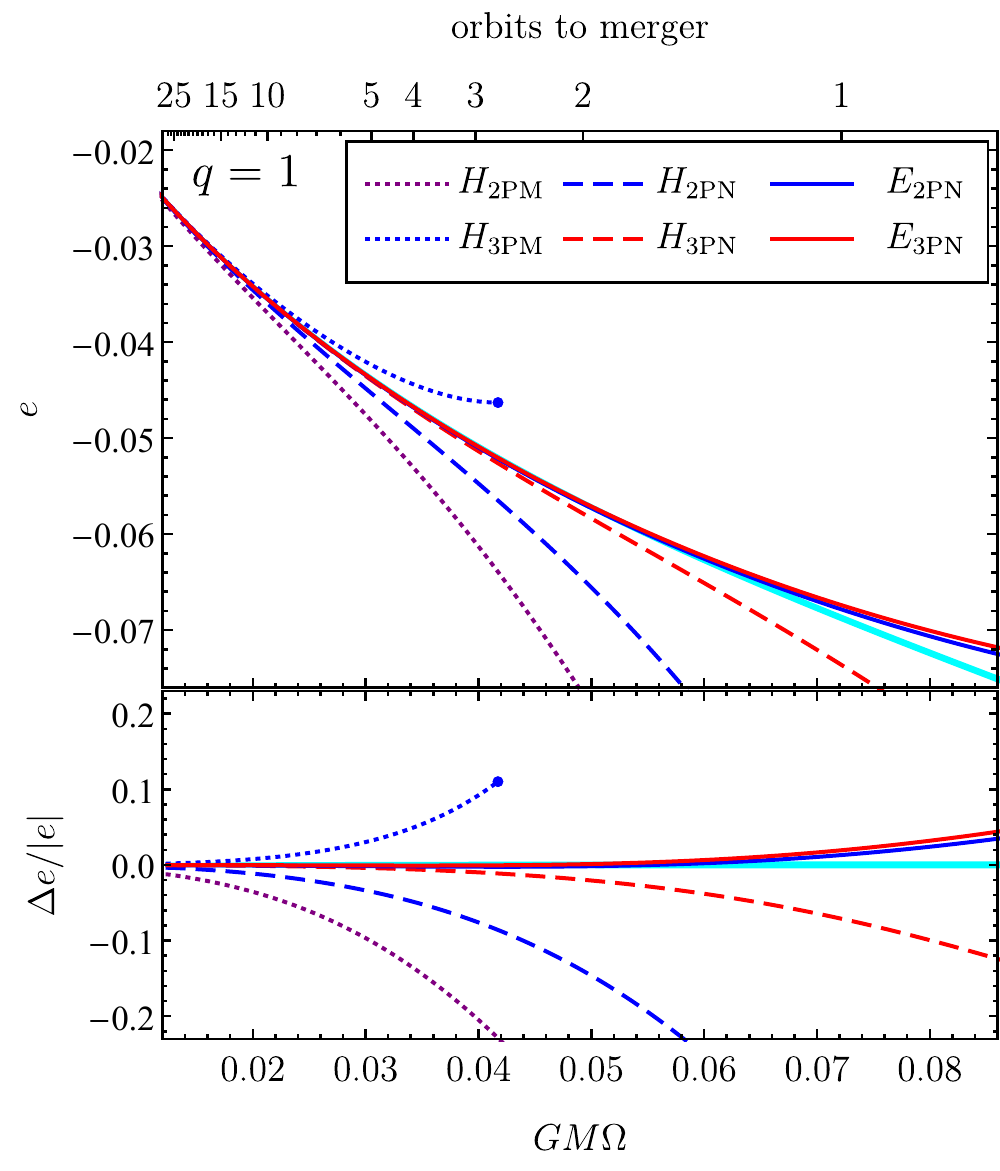}
	\caption{{\bf Energetics of PM and PN approximants versus analytic PN calculations.} As in previous figures, the dotted and dashed curves show the binding energy obtained numerically from the $H^{\rm EOB}_{m \rm PM}$ and $H^{\rm EOB}_{n \rm PN}$ Hamiltonians. The solid curves labeled ``$n$PN'' show the binding energy computed order-by-order in $GM\Omega$~\cite{Blanchet:2013haa}.
	\label{fig:vanPMvsPNq1}
	}
\end{figure}

In Fig.~\ref{fig:PMvsclassicEOB}, for mass ratio $q=1$, we
show how the $H^{\rm EOB,PS}_{m \rm PM}$ Hamiltonian compares with the 
(original) $H^{\rm EOB}_{n \rm PN}$ Hamiltonian, currently employed (after further improvements from NR simulations) 
in LIGO/Virgo searches and data analysis~\footnote{We note that the upper right panel of Fig. 4 in Ref.~\cite{Ossokine:2017dge} also shows a comparison between the binding energy from the EOB Hamiltonian and NR predictions. However, the agreement to NR differs from ours in Fig.~\ref{fig:PMvsclassicEOB}, because Ref.~\cite{Ossokine:2017dge} employs the EOB Hamiltonian where the potential for circular orbits has been resummed 
as suggested in Ref.~\cite{Barausse:2009xi}, and it computes the binding energy through a quasi-circular inspiral.}. We find that $H^{\rm EOB}_{n \rm PN}$
always leads to a binding energy that is closer to the NR one. Thus,
we find that insights on (and explicit resummations for) the EOB
Hamiltonian from current PM results are not yet sufficient to improve
the accuracy of quasi-circular inspiral waveforms for LIGO/Virgo data
analysis. This is not entirely surprising, because the currently known
3PM level only covers completely the 2PN level of the PN
approximation; there is much room (hope) for improvement coming from
4PM. The conclusion is that it will be very useful to
extend the knowledge of PM calculations to higher orders --- for
example at least 4PM, but even 5PM order. 

Before ending this section we remark that the comparison
results that we have illustrated  depend on several choices. 
First of all, we have decided to compare the binding energy
extracted from NR simulations to results obtained from an adiabatic
sequence of circular orbits, instead of the ones from the Hamilton equations 
with radiation-reaction force. To illustrate the impact of this choice we compare in 
Fig.~\ref{fig:evolq1} the binding energies of $H^{\rm EOB}_{3 \rm PN}$
and $H^{\rm EOB,PS}_{n \rm PN}$ obtained by evolving the Hamilton equations with 
a suitable radiation-reaction force 
(labeled ``inspiral'') and using an adiabatic sequence of circular orbits (labeled ``circular''). 
The difference is small early in the evolution and grows as the inspiral approaches the ISCO,
where we observe a typical difference in the binding energy of 5\% to
10\% (for $q=1$).

Lastly, Fig.~\ref{fig:vanPMvsPNq1} demonstrates the difference of calculating
$e(\Omega)$ numerically, treating the various approximants of the
Hamiltonian as exact, and analytically as an expansion in
$(GM\Omega)$. The plots show the results of calculating $e(\Omega)$
numerically from $m$PM and $n$PN Hamiltonians treated as ``exact'', and also
the curves from the analytically computed binding-energy $E_{n\rm PN}(\Omega)$ truncated at 2PN
(i.e., $(GM\Omega)^{6/3}$ with respect to leading term) and 3PN (i.e., $(GM\Omega)^{8/3}$) order (see Eq.~(232) in 
Ref.~\cite{Blanchet:2013haa}) (labeled $E_{n\rm PN}$). As already noticed 
in Ref.~\cite{Buonanno:2005xu}, the differences can be quite substantial. However, it is
worth re-emphasizing that if one calculates $e(\Omega)$ analytically
starting from either $H_{3\rm PM}$ or $H_{2\rm PN}$ one recovers the
2PN result exactly.

\section{Conclusions}
\label{sec:concl}

The study of the energetics conducted in this work, using currently available PM Hamiltonians 
up to third order, highlights two main points.  Firstly, the binding energy 
for circular orbits computed with the 3PM Hamiltonian of Ref.~\cite{Bern:2019nnu}
and the 3PM EOB Hamiltonian of Sec.~\ref{sec:EOB3PM} are closer to NR predictions than 
the ones computed at lower PM orders, especially for small
frequencies (or high angular momenta) (see Figs.~\ref{fig:energywPM} and \ref{fig:energyJPM}). 
This suggests that similar improvements can be made by pushing PM calculations to higher orders, leading to
 a more accurate modeling of the inspiral phase.

Secondly, we find that higher-order PM calculations of the conservative 
two-body dynamics would be needed to improve the agreement to NR and compete with (the
conservative part of) currently available waveform models used in LIGO/Virgo 
data analysis (see Figs.~\ref{fig:energymixed} and \ref{fig:PMvsclassicEOB}).
This is not surprising, since the 3PM order contains complete PN information only up to 2PN order, but also not obvious \textit{a priori}, since the 3PM approximation contains information not available in any of the PN expansions.

Furthermore, we have found that the PM EOB Hamiltonian of Sec.~\ref{sec:EOB3PM} (originally 
derived at 2PM order in Ref.~\cite{Damour:2017zjx}) gives 
good agreement against NR (and better agreement than the 3PM Hamiltonian of 
Ref.~\cite{Bern:2019nnu}), albeit not at the level of the PN 
EOB Hamiltonian~\cite{Buonanno:1998gg,Damour:2000we,Damour:2015isa} used to 
build waveform models for LIGO/Virgo data analysis (see Figs.~\ref{fig:energymixed} and ~\ref{fig:PMvsclassicEOB}). 
Relatedly, in Fig.~\ref{fig:energymixed} we have also shown the binding energy computed with a 3PM EOB Hamiltonian 
that we have derived in Appendix \ref{appendixB} and that differs from the one of Sec.~\ref{sec:EOB3PM} at 4PM order. 
Interestingly, we have found that such an alternative EOB Hamiltonian has much better agreement with NR than 
the one of Sec.~\ref{sec:EOB3PM} (e.g., confront the lower panels of Fig.~\ref{fig:energymixed} 
and Fig.~\ref{fig:PMvsclassicEOB}), reaching agreement similar to the EOB Hamiltonian 
employed to construct waveform models for LIGO/Virgo detectors (the latter would still do much better in the low frequency early inspiral not covered by the NR simulation). This rather 
encouraging result  motivates a more comprehensive study of EOB resummations of PM results. 

We recall that there are several caveats that underlie our investigation. 
To begin with, we have chosen to work in the circular-orbit approximation, rather than 
incorporating radiation-reaction effects and evolving the two-body system along an 
inspiraling orbit. This choice was dictated by the desire of clearly singling out 
the contribution coming from the conservative dynamics in the PM, PN and EOB descriptions. 
We have also decided to treat perturbative PM and PN Hamiltonians as exact when computing 
the binding energy. The effect of each of these choices has been illustrated 
in Figs.~\ref{fig:evolq1} and \ref{fig:vanPMvsPNq1}.

It is relevant to extend the above comparisons to quasi-circular inspiraling orbits (i.e., 
including  radiation-reaction effects), and we plan to do so in the near future.  
It would also be very interesting to perform the comparisons for scattering/unbound orbits, 
i.e. a setting closer to the natural domain of the PM approximation. It would be desirable,
for instance, to compare observables like the scattering angle against
NR simulations, as initiated in Ref.~\cite{Damour:2014afa}. Here, in a regime of high 
impact velocity and large impact parameter, the PM Hamiltonians are expected to behave better
than pure PN ones.

\section*{Acknowledgments}
We thank the authors of Ref.~\cite{Bern:2019nnu} for discussions and for sharing with us the results of their 3PM Hamiltonian 
while finalizing their paper. It is also a pleasure to thank Sergei Ossokine for making available to us the NR data for the binding energy and Chris Kavanagh for useful discussions.  

MvdM was supported by European Union's Horizon 2020 research and innovation program under grant agreement No.~705229.

\appendix

\section{Effective-one-body Hamiltonian at 3PM order augmented by 3PN and 4PN information}\label{PM4PN}
\label{appendixA}

The 3PM EOB Hamiltonian given in Sec.~\ref{sec:EOB3PM}, like the BCRSSZ Hamiltonian from which it was derived, encodes the complete conservative dynamics for generic orbits up to 2PN order, as well as partial information at higher PN orders.  Here we discuss how further information from 3PN and 4PN calculations can be added to the 3PM EOB Hamiltonian, focusing on the case of bound (near-circular) orbits.

We recall that the 4PN Hamiltonian as applicable to generic orbits~\cite{Damour:2014jta,Damour:2016abl,Bernard:2016wrg} is \emph{not} a usual local-in-time Hamiltonian, i.e., not a function of instantaneous position and momentum; rather, it contains a contribution which is a nonlocal-in-time functional of the phase-space trajectory---the so-called ``tail'' term.  In Ref.~\cite{Damour:2015isa}, an EOB transcription of the generic nonlocal-in-time 4PN Hamiltonian is evaluated as a usual local-in-time Hamiltonian by implementing an expansion about the circular-orbit limit, i.e., and expansion in small eccentricity or equivalently in small $\hat{p}_r$.  The result for the 4PN (reduced) effective Hamiltonian takes the form
\be
(\hat H^\mr{eff}_\mr{4PN})^2=A\Big(1+l^2u^2+A\,\bar D\,\hat p_r^2+\hat Q\Big),
\ee  
where we recall that 
\be
l=\frac{L}{GM\mu},
\quad
\hat p_r=\frac{p_r}{\mu},
\quad
u=\frac{GM}{r},
\ee
with the potentials $A(u,\nu)$, $\bar D(u,\nu)$, and $\hat Q(u,p_r,\nu)$ at 4PN order given by
\begin{alignat}{3}
A&=1-2u+2\nu u^3+a_4u^4+(a_{5,\mr c}+a_{5,\ln}\ln u)u^5,
\nnm\\\nnm
\bar D&=1+6\nu u^2+\bar d_3 u^3+(\bar d_{4,\mr c}+\bar d_{4,\ln}\ln u)u^4,
\\\nnm
\hat Q&=q_{42} \hat p_r^4u^2+(q_{43,\mr c}+q_{43,\ln}\ln u)\hat p_r^4u^3
\\
&\quad+(q_{62,\mr c}+q_{62,\ln}\ln u)\hat p_r^6u^2+\mc {\cal O}(\nu \hat p_r^8 u ).
\label{DJSpotentials}
\end{alignat}
The coefficients up to 2PN order have been written explicitly here, while the 3PN coefficients ($a_4$, $\bar d_3$, $q_{42}$) and 4PN coefficients are functions only of $\nu$ and are given in Eqs.~(8.1) of Ref.~\cite{Damour:2015isa}.  The $A$ and $\bar D$ potentials are complete up to 4PN order, while the $\hat Q$ potential is given at 4PN order as an expansion in $\hat p_r$ (small-eccentricity expansion) up to ${\cal O}(\hat p_r^6)$, and thus is valid only in the near-circular-orbit regime.

One way to add the 3PN and 4PN information to the 3PM EOB Hamiltonian derived in Sec.~\ref{sec:EOB3PM} is to find a canonical transformation which brings the above 4PN Hamiltonian~\cite{Damour:2015isa} into a form matching (the PN expansion of) the following 3PM+4PN ansatz.  As a natural generalization of the 2PM+3PN ansatz in Ref.~\cite{Damour:2017zjx}, we consider a post-Schwarzschild (reduced) effective Hamiltonian of the form
\be
\Big[\hat H^\mr{eff,PS}(u,\hat p_r,l)\Big]^2=
\hat H_\mr S^2
+(1-2u)\hat Q^\mr{PS}(u,\hat H_\mr S,\nu),
\label{HPSapp}
\ee
where $\hat H_\mr S=\sqrt{1-2u}\sqrt{1+l^2u^2+(1-2u)\hat p_r^2}$ is the reduced Schwarzschild Hamiltonian.  Imposing this form, with a dependence on $\hat p_r$ and $l$ only through $\hat H_S$, is seen to fix a unique phase-space gauge choice.  The resultant potential $\hat Q^\mr{PS}$ can be written at 3PM+4PN order as
\begin{alignat}{3}
\label{QPS}
\hat Q^\mr{PS}&=u^2q_\mr{2PM}(\hat H_\mr S,\nu)
+u^3q_\mr{3PM}(\hat H_\mr S,\nu)
\\\nnm
&\quad
+\Delta_\mr{3PN}(u,\hat H_\mr S,\nu)
+\Delta_\mr{4PN}(u,\hat H_\mr S,\nu)+\mc O(\mr{5PN}).
\end{alignat}
This differs from Eq.~(\ref{3PMeffH}) by the addition of the $\Delta$ terms, which are given as expansions in the two PN small parameters $u$ and $\hat H_S^2-1$ (each $\mc O(1/c^2)$), at the orders needed to find a unique match to the 4PN EOB Hamiltonian of Ref.~\cite{Damour:2015isa}.  At 3PN order, for generic orbits, we need only a single 4PM term [given by Eq.~(6.3) in Ref.~\cite{Damour:2017zjx}], at zeroth order in $\hat H_\mr S^2-1$,
\be
\Delta_\mr{3PN}=\bigg(\frac{175}{3}\nu-\frac{41\pi^2}{32}\nu-\frac{7}{2}\nu^2\bigg)u^4.
\ee
At 4PN order, to match the near-circular-orbit expansion of the potential $\hat Q$ in Eq.~\eqref{DJSpotentials} up to $\mc O(\hat p_r^6)$, we must have 
\begin{alignat}{3}
\Delta_\mr{4PN}&=\sum_{n=2}^5\alpha_{4n}u^n(\hat H_\mr S^2-1)^{5-n}
\nnm\\
&\quad+\Big(\alpha_{44,\ln}u^4(\hat H_\mr S^2-1)+\alpha_{45,\ln}u^5\Big)\ln u,
\end{alignat}
where the $\alpha$'s are functions only of $\nu$.  (The $n=2,3$ terms here arise solely from the nonlocal tail integral, while the $n=4,5$ and $\ln$ terms include local and tail contributions.)  Implementing the canonical transformation from the 4PN EOB Hamiltonian 
of Ref.~\cite{Damour:2015isa}, we find the coefficients
\begin{alignat}{3}
\alpha_{42}&=
\bigg({-}\frac{1027 }{12}-\frac{147432}{5}   \ln2
\nnm\\
&\quad
+\frac{1399437}{160}   \ln3
+\frac{1953125}{288}  \ln5\bigg)\nu\,,
\\\nnm
\alpha_{43}&=
\bigg({-}\frac{78917 }{300}-\frac{14099512}{225}  \ln 2
\\
&\quad
+\frac{14336271}{800} \ln
3+\frac{4296875}{288} \ln 5\bigg)\nu\,,
\\\nnm
\alpha_{44}&=\bigg({-}\frac{43807}{225}
+\frac{296 \gamma_{\text{E}} }{15}
-\frac{33601 \pi ^2}{6144}
\\\nnm
&\quad-\frac{9771016 }{225}\ln 2
+\frac{1182681}{100} \ln 3
+\frac{390625}{36} \ln 5
\bigg)\nu 
\\
&\quad
+\bigg({-}\frac{405}{4}+\frac{123}{54}\pi^{2}\bigg)\nu^{2}+\frac{13}{2}\nu^{3}\,,
\\\nnm
\alpha_{45}&=\bigg({-}\frac{34499}{1800}+\frac{136 }{3} \gamma_{\text{E}}-\frac{29917 }{6144}\pi ^2
\\\nnm
&\quad-\frac{254936}{25}\ln 2+\frac{1061181}{400} \ln 3+\frac{390625 }{144}\ln 5\bigg)\nu 				
\nonumber \\
&\quad+\bigg({-}\frac{2387}{24}+\frac{205 }{64}\pi ^2\bigg) \nu ^2+\frac{9 }{4}\nu ^3\,,
\end{alignat}
and
\be
\alpha_{44,\ln}=\frac{148}{15}\nu\,,
\qquad
\alpha_{45,\ln}=\frac{68}{3}\nu\,.
\ee
It is important to note, again, that the form of the effective Hamiltonian in Eq.~(\ref{HPSapp}) and (\ref{QPS}) 
(notably the 4PN term $\Delta_\mr{4PN}$ in Eq.~(\ref{QPS}), is only valid for bound orbits
in the small-eccentricity expansion (around the circular-orbit case)\footnote{If we
included the next term in the small-eccentricity expansion (i.e.,
a $\mc {\cal O}(\nu \hat p_r^8 u )$ term in \eqref{DJSpotentials}),
then all coefficients $\alpha_{4n}$ are modified and we need to
introduce an additional coefficient for $n=1$. This illustrates that our PS transcription of the local EOB Hamiltonian at 4PN order is not optimal when we have only a finite number of terms in the $\hat p_r$ expansion.  A local 4PN-3PM EOB Hamiltonian 
for scattering orbits could be obtained by matching $\Delta_\mr{4PN}$ to the 4PN scattering angle calculated 
in Ref.~\cite{Bini:2017wfr} in the large-eccentricity limit.}.

The two-body Hamiltonian in the EOB framework is then obtained by inserting the effective Hamiltonian (\ref{HPSapp}) in 
Eq.~(\ref{3PMEOBH}), thus obtaining $H^\mr{EOB,PS}_\mr{3PM+4PN}$, or $H^\mr{EOB,PS}_\mr{3PM+3PN}$, if we include $\Delta_\mr{3PM}$, but drop $\Delta_\mr{4PN}$.  The Hamiltonian $H^\mr{EOB,PS}_\mr{4PN}$ is obtained by expanding $q_\mr{2PM}$ to $\mc O(\hat H_\mr S^2-1)^3$ and $q_\mr{3PM}$ to $\mc O(\hat H_\mr S^2-1)^2$, while for $H^\mr{EOB,PS}_\mr{3PN}$ we keep one less order of $\hat H_\mr S^2-1$ for each $q$ and drop $\Delta_\mr{4PN}$.

\section{Alternative effective-one-body Hamiltonian at 3PM order for circular orbits}
\label{appendixB}

One straightforward alternative form for a 3PM EOB Hamiltonian can be obtained simply by fully expanding the right-hand-side of Eq.~\eqref{HPSapp} in $G$, to $\mc O(G^3)$.  Here we explicitly state the result of this expansion evaluated at $p_r=0$, which determines the circular-orbit binding-energy approximants:
\begin{align}
(\hat H_\mr{3PM}^{\mr{eff},\widetilde{\mr{PS}}})^2|_{p_r=0}&=(1-2u)(1+l^2u^2) \nonumber \\
&\quad+u^2 \tilde q_\mr{2PM}(\gamma_0,\nu) \nonumber \\
&\quad+u^3\tilde q_\mr{3PM}(\gamma_0,\nu)
+\mc O(G^4)\,,
\end{align}
where
$
\gamma_0=\sqrt{1+l^2u^2}
$
is the (circular) effective Hamiltonian at zeroth order in $G$, with 
\begin{subequations}
\begin{align}
\tilde q_\mr{2PM}(\gamma_0,\nu)&=q_\mr{2PM}(\gamma_0,\nu),
\\
\tilde q_\mr{3PM}(\gamma_0,\nu)&=q_\mr{3PM}(\gamma_0,\nu)
\nonumber \\
&\quad
-\frac{3\nu\gamma_0(5\gamma_0^2-1)}{2\Gamma_0^3}-3(10\gamma_0^2-1)\Big(1-\frac{1}{\Gamma_0}\Big)\,,
\end{align}
\end{subequations}
where the functions 
$q_\mr{2PM}(\hat H_\mr S,\nu)$ and $q_\mr{3PM}(\hat H_\mr S,\nu)$ are given by Eqs.~\eqref{q2q3}, and $\Gamma_0=\sqrt{1+2\nu(\gamma_0-1)}$.

\raggedright
\bibliography{refs}

\end{document}